\documentclass[english,12pt]{article}
\usepackage{array}
\usepackage{graphicx}
\usepackage{amssymb}
\usepackage{amsmath}
\usepackage{multirow}
\usepackage{prettyref}
\usepackage{babel}
\usepackage{units}
\usepackage[latin1]{inputenc}
\usepackage{amsfonts}
\usepackage{amssymb}
\usepackage{babel}
\usepackage{cite}
\def\@fmsl@sh#1#2#3{\m@th\ooalign{$\hfil#1\mkern#2/\hfil$\crcr$#1#3$}}
 \def\eq#1\en{\begin{equation}#1\end{equation}}
\def\s[#1,#2]{[#1\stackrel{\star}{,}#2]}
\def\sx[#1,#2]{[#1\stackrel{\star_{x}}{,}#2]}

\newcommand{\nc}{\newcommand}
\nc{\beq}{\begin{equation}}
\nc{\eeq}{\end{equation}}
\nc{\beqa}{\begin{eqnarray}}
\nc{\eeqa}{\end{eqnarray}}

\def\bc{\begin{center}}
\def\ec{\end{center}}

\def\gsim{\mathrel{\mathpalette\atversim>}}

\def\bc{\begin{center}}
\def\ec{\end{center}}

\def\gsim{\mathrel{\rlap{\lower4pt\hbox{\hskip1pt$\sim$}}

    \raise1pt\hbox{$>$}}}       

\def\gsim{\mathrel{\rlap{\lower4pt\hbox{\hskip1pt$\sim$}}
    \raise1pt\hbox{$>$}}}       

\begin{document}
\makeatletter
\def\fmslash{\@ifnextchar[{\fmsl@sh}{\fmsl@sh[0mu]}}
\def\fmsl@sh[#1]#2{%
  \mathchoice
    {\@fmsl@sh\displaystyle{#1}{#2}}%
    {\@fmsl@sh\textstyle{#1}{#2}}%
    {\@fmsl@sh\scriptstyle{#1}{#2}}%
    {\@fmsl@sh\scriptscriptstyle{#1}{#2}}}
\def\@fmsl@sh#1#2#3{\m@th\ooalign{$\hfil#1\mkern#2/\hfil$\crcr$#1#3$}}
\makeatother

\thispagestyle{empty}
\begin{titlepage}
\boldmath
\begin{center}
  \Large {\bf Virtual Black Holes, Remnants and the Information Paradox}
    \end{center}
\unboldmath
\vspace{0.2cm}
\begin{center}
{  {\large Xavier Calmet}\footnote{x.calmet@sussex.ac.uk}}
 \end{center}
\begin{center}
{\sl Physics $\&$ Astronomy, 
University of Sussex,   Falmer, Brighton, BN1 9QH, United Kingdom 
}
\end{center}
\vspace{5cm}
\begin{abstract}
\noindent
We revisit the question of the contributions of Planckian quantum black holes in general and of remnants in particular to low energy physics observables. As long as quantum gravity preserves the symmetries of the low energy effective field theory, we find that the bounds on the number of quantum black holes or remnants are very weak. Typically we rule out using data on the anomalous magnetic moment of the muon  that there are more than $10^{32}$ quantum black holes coupled to the standard model particles gravitationally. Remnants thus remain a viable option as a solution to the information paradox of black holes. 
\end{abstract}  
\end{titlepage}



\newpage

Black holes come in a wide range of masses from supermassive black holes at the center of galaxies to Planck-size quantum black holes. While astrophysical black holes have been observed, quantum black holes are much more speculative but to a certain extend also much more interesting since a proper description of their physical properties requires to  understand general relativity in the quantum regime. 

A stationary black hole is a very simple object which can be fully described by only three quantities namely its mass, its angular momentum and its electric charge. This is a consequence of the famous no-hair theorem \cite{Wald:1984rg}. Because black holes are characterised by a few quantum numbers, it is tempting to treat them as elementary particles and thus to include them in the Hilbert space, at least for the lightest of these objects.

The mass of a black hole is linked to its temperature. If the mass of the black hole is much larger than the Planck scale $M_P$, it is a classical object and it has a well defined temperature. The semi-classical region starts between 5 and 20 times the Planck scale \cite{Meade:2007sz}.  Semi-classical black holes are also thermal objects. On the other hand, black holes with masses of the order of the Planck scale are non-thermal objects \cite{Calmet:2008dg}. We shall call these Planckian objects quantum black holes. A thermal black hole will decay via Hawking radiation and thus couples effectively to many degrees of freedom. The decay of a non-thermal black hole is not well described by Hawking radiation. Rather than decaying to many degrees of freedom, one expects that it will only decay to a few particles only, typically two because this object is non-thermal. 

The production of black holes in the high energy collision of elementary particles can be modeled by the collision of shockwaves. In the limit of the center of mass $E_{CM}$ going to infinity, Penrose \cite{Penrose}  and independently Eardley and Giddings \cite{Eardley:2002re} have shown that even when the impact parameter is non zero a classical black hole ($M_{BH} \sim E_{CM}  \gg M_P$) will form. They were able to prove the formation of a closed trapped surface. Their result justifies using the geometrical cross to calculate the cross section production of black holes in the high energy collisions of two particles. It is given by
\begin{eqnarray} \label{eq1}
\sigma = \pi r_S^2\theta(s-M_{BH}^2)\sim \frac{s}{M_{BH}^4} \theta(s-M_{BH}^2),
\end{eqnarray}
 where $s=E_{CM}^2$ is the center of mass squared, $r_S$ the Schwarzschild radius and $\theta$ is the Heaviside step function.  The step function implies a threshold for black hole formation. The work of Eardley and Giddings can be extrapolated into the semi-classical regime using path integral methods \cite{Hsu:2002bd}. A final leap of faith leads to an extrapolation into the full quantum regime. It is usually assumed that the geometrical cross section holds for Planck size black holes as well. This has interesting consequences as we shall see shortly.

It is often argued that Planck size black holes may impact low energy measurements because of the large multiplicity of states. This is particularly true if one thinks of Planck size black holes as remnants which could resolve the information paradox of black holes, see e.g. for a review \cite{Strominger:1994tn}, by storing the information within the volume in their Schwarzschild radius. 

Our first observation is that the on-shell production of the lightest possible black holes, i.e. Planckian quantum black holes, if we accept the geometrical cross section, would require doing collisions at the Planck scale which is conservatively taken to be of the order of $10^{19}$ GeV since there is a step function in energy which implies an energy threshold. We have never probed physics beyond the few TeV region directly at colliders and cosmic ray collisions have center of mass energies of a few 100 TeV.  Unless we live in a world with large extra-dimensions \cite{ArkaniHamed:1998rs,Randall:1999ee} or with large hidden sector of hidden particles \cite{Calmet:2008tn}, there is no reason to expect to produce on-shell Planckian quantum black holes in low energy experiments since the center of mass energy of such collisions is below the production threshold according to the geometrical cross section. Direct production thus cannot probe the existence of Planckian quantum black holes or remnants since we have to take $M_{BH} \sim M_P$ in Eq. (\ref{eq1}) .

If one considers quantum field theoretical corrections to particle physics processes, the situation is different. Let us consider the contribution of quantum black holes into loops, i.e. virtual quantum black holes. For definiteness let us consider a single spin-0 black hole with mass $M_{BH}$.  If we close a loop with a massive scalar field of mass $M_{BH}$, one expects contributions of the type
\begin{eqnarray}
I=\int_0^\Lambda d^4p \frac{1}{p^2-M_{BH}^2+ i\epsilon}
\end{eqnarray}
where $\Lambda$ is some ultra-violet cutoff. Such integrals behave as $\Lambda^4/M_{BH}^2$ for momenta much smaller than $M_{BH}$. The cutoff $\Lambda$ is much smaller than $M_{BH}$ since we are looking at low energy experiments. Heavy particles decouple from the low energy effective theory as naively expected. When one calculates the anomalous magnetic moment of the muon, one need not worry about very high energy embeddings of the standard model such as grand unified theories. One probes, as we shall see shortly, at most the few TeV region if new physics respects chirality or the $10^7$ GeV region if it does not. As long as a high energy theory does not violate symmetries of the low energy effective theory, one expects its particles to decouple from the low energy regime.  Note that one may worry that since  quantum black holes are genuinely quantum gravitational objects and hence one might not be cannot calculate their production rate using an effective field theory approach. However, as long as we do physics well below the Planck scale, their should exists a quantum field description of such states and they can be classified according to representations of the Lorentz group. This does not imply that the internal geometry of the quantum black hole cannot be very large and cannot contain lots of information as we shall assume in the sequel when we discuss the resolution of the information paradox by remnants.

The situation for quantum black holes is different since the spectrum of quantum gravity contains potentially a large number of states. If we sum over the number $N$ of scalar fields with masses $M_{BH,i}$, these contributions can be very large and potentially impact in a sizable way low energy observables. In the case of a continuous mass spectrum however, the sum is replaced by an integral over the mass spectrum of the black holes. We have
\begin{eqnarray}
I=\int_{M_{BH,l}}^{M_{BH,h}} \frac{\Lambda^4}{M^2} \rho(M) dM
\end{eqnarray}
where $\rho(M_{BH})$ is the black hole mass density, $M_{BH,l}$ is the lightest black hole mass, while ${M_{BH,h}}$ is the heaviest mass a black hole can have. For a single black hole,  $\rho(M_{BH})=\delta(M-M_{BH})$ while for a continuous mass spectrum, one has $\rho(M)=N M^{-1}$ where $N$ is the number of black states which leads to
\begin{eqnarray}
I_{continuous}=\int_{M_{BH,l}}^{M_{BH,h}} \frac{\Lambda^4}{M^2} \rho(M) dM\sim \frac{\Lambda^4(M_{BH,h}^2-M_{BH,l}^2)}{M_{BH,h}^2 M_{BH,l}^2}N.
\end{eqnarray}
Here $N$ is the number of black holes states between $M_{BH,h}$ and $M_{BH,l}$, which is indeed infinite for a continuous mass distribution.  Furthermore, in the case of remnants as a solution to the information paradox, it is argued that their might be large multiplicity factor ${\cal M}$ arising from a sum over all the possible quantum numbers of the black holes contributing in the loop.  This is the standard argument against the resolution of the black hole information paradox based on remnants \cite{Giddings:1992hh}. It would apply as well to quantum black holes predicted by models of low scale quantum gravity.  Our results are thus useful also independently of question of the information paradox of black holes.

The aforementioned work on the production of black holes in the collisions of particles at very high energy can help us to identify reasonable values for $M_{BH,h}$ and $M_{BH,l}$. The lightest black hole produced cannot have a mass below $M_P$, we shall thus identify $M_{BH,l}\sim M_P$. On the other hand, we know that black holes with mass 5 to 20 times $M_P$ are semi-classical objects. Let us briefly discuss the criterion given in \cite{Meade:2007sz} for a small black hole to be thermal. If one considers the high energy collision of two particles, one could require that the Compton wavelength of the colliding particle of energy $E_{CM}/2$ lies within the Schwarzschild radius for a black hole of given energy $E_{CM}$. If we do that, for example, for a 4-dimensional black holes, we find that the first semi-classical black hole as a mass of 12.6 $M_P$. A weaker criteria would be to require $r_S > 1/E_{CM}$, in which case the semi-classical regime would start at 3.5 $M_P$. In any case, semi-classical black holes, from an effective theory point of view, are states that couple to many particles (Hawking radiation photons). In contrast to  quantum black holes, they are thus unlike particles which typically only couple to a few other particles. It thus does not make much sense to include these objects in the Hilbert space and we should thus identify $M_{BH,h}$ with 5-20 $M_P$. The contribution of quantum black holes to the loop integral discussed above is thus of the order of 
\begin{eqnarray}
I_{continuous}=  \frac{\Lambda^4}{M_P^2}N {\cal M}  
\end{eqnarray}
Since $\Lambda \ll M_P$ as we are interested in low energy experiments, the number of state $N$ the potentially large multiplicity ${\cal M}$ are the source of potential large contributions to low energy physics observables. 

An obvious solution to the large (actually infinite) factor $N$ is that the spectrum of quantum black holes with masses up to 5-20 $M_P$ is quantized. This is perfectly reasonable as we have strong arguments in favor of a quantization of space-time in terms of the Planck scale \cite{Calmet:2007vb,Dvali:2011nh}. If we assume that the mass spectrum is quantized in terms of $M_P$ then $N=5-20$ and is not a large factor. 
 
Let us now discuss how large ${\cal M}$ might be. Its value depends on whether quantum black holes have hair or not. If we naively extrapolate from classical objects, one would expect the no-hair theorem to hold. In the case of remnants one could argue that the information is contained inside the black hole horizon but that for an observer outside the black hole, the black hole is still described in terms of very few quantities, namely its mass, its angular momentum and its electric charge. In that case, the multiplicity factor ${\cal M}$ is small and the contribution of quantum black holes to low energy observables is negligible. The following thought experiment shows that in all likelihood quantum black holes are slightly more complicated than their classical counterparts. If we think of the creation of a quantum black holes in the collision of two colored particles, we have to accept that either the black hole is not formed or that the quantum black hole will carry the color charges of the particles which created it.  Quantum numbers corresponding to gauged quantities must be conserved. However, in that case we do not expect ${\cal M}$ to be large, it will merely be a group theoretical factor. Such factors are usually of order unity.  While the no-hair theorem probably cannot be valid for quantum black holes if they exist, we do not expect that there will be a multitude of new quantum numbers carried by the black holes, merely the quantum numbers corresponding to the gauge groups of the standard model of particle physics.  

Even though two remnants may contain different information inside their Schwarzschild radius, if their quantum numbers observed by an outside observer are the same, they should be treated as only one state of the Hilbert space and there will not be a large multiplicity of states from the low energy effective theory point of view. Let us stress that this is the main point of disagreement with the calculations presented in \cite{Giddings:1992hh} where it is assumed that each remnant carries a different and observable quantum number and that one should sum over all these states. We  wish to emphasize that this is an assumption and not a necessity. Our assumption is actually supported by a recent study   \cite{Smolin:2005tz} which argues that a very small object can lock lots of information, although the real amount of information of the remnant as well as that of Hawking radiation are not so large until the last stage of evaporation.

One may worry of consequences of this assumption for the thermodynamics of black holes. In particular, it has been claimed that entropy bounds strongly restricts the information capacity of  black hole remnants, so that they cannot serve to resolve the information paradox \cite{Bekenstein:1993dz,Marolf:2003wu}. However, we are considering objects with masses close to the Planck mass, they are thus highly non-thermal objects. Their entropy is not necessarily well defined as is their temperature. However, in specific quantum black hole models, see e..g \cite{Nicolini:2012eu}, such quantities can be defined. However, without a complete theory of quantum gravity, it is difficult to exclude remnants based on this argument. 
 
The resolution of this divergence of viewpoints cannot be resolved without a full theory of quantum gravity.  Let us emphasize further that it is not necessary for the remnants to carry ``observable" quantum numbers to resolve the information paradox. The information paradox is due to Hawking radiation which is a thermal radiation that cannot carry information. In our case, when the mass of the black hole reaches the Planck scale, Hawking radiation stops and a remnant remains that carries the information inside its Schwarzschild radius. This information is not observable to an observer outside the Schwarzschild radius, but it is not lost.

 \begin{figure}
\centerline{\includegraphics[height=4cm]{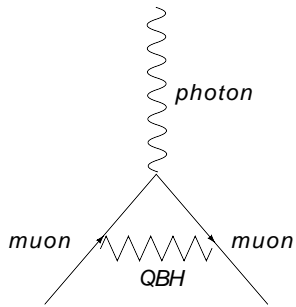}} \label{fig1}
\caption{
{\it Contribution of a quantum black hole (QBH) to the muon anomalous magnetic moment.} 
}
\end{figure}

We now show that the number of quantum black hole states is not  strongly constrained by low energy experiments.  One of the most precise experiments done to date is that of the measurement of the anomalous magnetic moment of the muon. If gravity respects chiral symmetry as perturbative quantum gravity indicates, Quantum black holes will typically lead to dimension 6 operators of the type \cite{Calmet:2011ta}
\begin{eqnarray} \label{N}
N  \frac{e}{2} \frac{m_\mu}{ 16 \pi^2 \bar M_P^2} \bar \psi \sigma_{\mu\nu} \psi F^{\mu\nu} 
\end{eqnarray}
where  $e$ is the electron charge, $N$ is the number of quantum black holes propagating in the diagram depicted in Fig. 1, $\bar M_P$ is the reduced Planck mass, $m_\mu$ is the muon mass, $\psi$ its wavefunction and $F^{\mu\nu}$ the electromagnetic field strength tensor. The generic bound on the scale of new physics $\Lambda_{NP}$ which suppresses a dimension six operator $(e/2 \times  m_\mu/ \Lambda_{NP}^2) \bar \psi \sigma_{\mu\nu} \psi F^{\mu\nu}$  is of the order of 2 TeV\cite{Calmet:2001dc}.  We can thus use this result to set a bound on $N$ which appears in Eq. (\ref{N}). We find $N<16 \pi^2 M_P^2/\Lambda_{NP}^2 \sim 10^{32}$ which is a very weak bound. We thus see that unless there is truly an infinite number of quantum black holes states, they cannot impact low energy observables in a sizeable manner.

The bound is slightly tighter if chirality is violated by quantum gravity at the non-perturbative level, one expects low energy effective operators of the type
\begin{eqnarray}
N  \frac{e}{2} \frac{1}{\bar M_P}  \bar \psi \sigma_{\mu\nu} \psi F^{\mu\nu}. 
\end{eqnarray}
Note that perturbative effects cannot violate chirality, if such an effect happens it is at the non-perturbative level and we thus do not include the factor $16 \pi^2$ in the denominator. The bound on the scale of new physics suppressing the operator $(e/2 \times  1/ \Lambda_{NP}) \bar \psi \sigma_{\mu\nu} \psi F^{\mu\nu}$ is of the order of  $2.5 \times 10^7$ GeV \cite{Calmet:2001dc}. We thus get a bound on $N$ of the order of $10^{11}$.

We have shown that the bounds on the number of quantum black holes (or remnants) interacting with low energy particles are rather weak unless some low energy symmetry is violated by quantum gravity. There is thus no reason, from a low energy effective theory point of view to rule out Planck size quantum black holes or remnants. Remnants are thus an acceptable solution to the black hole information paradox.

{\it Acknowledgments:}
This work is supported in part by the European Cooperation in Science and Technology (COST) action MP0905 ``Black Holes in a Violent  Universe" and by the Science and Technology Facilities Council (grant number  ST/L000504/1).


\bigskip{}

\end{document}